# STABILIZATION OF SPATIAL SOLITONS BY GAIN DIFFUSION

K.Staliunas
Physikalisch Technische Bundesanstalt, 38116 Braunschweig, Germany
tel: +49-531-5924429,  Fax: +49-531-5924423,  E-mail: Kestutis.Staliunas@PTB.DE

**Abstract**

It is shown, that diffusion of saturated gain (e.g. diffusion of population inversion in lasers) causes, and/or enhances a modulational instability of generated light field. In case of a subcritical system (e.g. a laser with saturable absorber) the enhancement of the modulational instability stabilizes spatial solitons. These predictions are made for general nonlinear optical systems, and are illustrated by numerical simulation of lasers with saturable absorber.

### I. Introduction

A well known transverse pattern formation mechanism in nonlinear optical resonators is off-resonance excitation. If the central frequency of a gain line of the laser $\omega_A$ is larger than the resonator resonance frequency $\omega_R$, then the excess of frequency $\Delta\omega = \omega_A - \omega_R$ causes transverse (spatial) modulation of laser field, with characteristic transverse wavenumber $\mathbf{k}$. This obeys a dispersion relation $a|\mathbf{k}|^2 = \Delta\omega$, where $a$ is the diffraction coefficient of the resonator. The off-resonance excitation mechanism is responsible for extended patterns (tilted waves, stripes, square vortex lattices, hexagons, e.a.) not only in lasers but also in other nonlinear optical resonators, like externally injected nonlinear resonators [1,2], photorefractive oscillators [3], or degenerate and nondegenerate optical parametric oscillators (DOPO and OPO) [4,5]. In case of subcritical [6,7] or nearly subcritical [8] nonlinear optical systems this off-resonance excitation mechanism is at the root of the localized pattern formation (spatial solitons or spatial localized structures: LS).

In this way, the spatial modulation of the fields manifests itself in extended or localized patterns. Such modulational instability is usually caused by geometric effects, namely due to interplay between the detuning of the resonator and the diffraction of the optical field. Other possible nonlocalities, such as diffusion of the population inversion in lasers, are often assumed to play only a secondary role. Either they are assumed to be negligibly small (like diffusion of population inversion in many types of lasers), or to counteract the modulational instability. Intuitively one would suppose, that diffusion in gain material (e.g. diffusion of population inversion in gas lasers, diffusion of free charge carriers in semiconductor lasers, *et cetera*) should reduce spatial inhomogeneities of the emitted optical field. As a consequence the gain diffusion would reduce or suppress a modulational instability, and might thus destroy extended or localized structures that exist for zero gain diffusion.

Some studies of patterns in nonlinear optics have been performed taking into account diffusion of population inversion [9-11]. In lasers weak diffusion of population inversion appears not to influence pattern forming processes [9]. For subcritical semiconductor lasers the diffusion of free charge carriers is shown to play a destructive role: the existence range of the spatial localized structures is reduced here by carrier diffusion [10,11]. The semiconductor lasers studied in [10,11] are, however, a particular system: the free charge carriers there are responsible not only for the gain, but also for change of refractive index of the medium. Thus the effects of diffusion of the charge carriers can not be interpreted solely by diffusion of gain as studied here.

We show here, that the diffusion of saturated gain enhances the spatial modulation of the optical fields. In case of a supercritical (monostable) laser the gain diffusion can enhance the spatial modulation, but can not cause a modulational instability. In the case of a subcritical (bistable) system the gain diffusion can not only enhance the spatial modulation, but can also cause the modulational instability. The enhancement of modulation supports the LSs (*i.e.* increases their stability range) in subcritical systems.

A general case of supercritical and of subcritical nonlinear optical systems is analyzed in this paper. Bright spatial LS in such systems can be interpreted in two ways: First, if a homogeneous solution corresponding to one of two bistability branches which is modulationally unstable, then the LS can occur as a homoclinic connection between a stable homogeneous state and the modulated state (stripe, hexagon), as shown by Fauve&Thual in [6]. The LS in this interpretation is a single isolated spot of a stripe (in 1D case) pattern, or of a hexagon pattern (in 2D). The background solution (solution far away from the LS) corresponds to the stable solution branch in this interpretation. For bright LSs the upper branch is usually modulationally unstable, and the radiation corresponding to the stable lower branch constitutes the background. For dark LS - *vice versa*.

When both the upper and lower branches are modulationally stable the LSs can be interpreted as homoclinic connection between the homogeneous states, as shown by Rosanov [7]. The spatial domain corresponding to one solution branch can contract to a finite minimum size, which is stabilized due to the interaction between domain boundaries. The bright LSs in this interpretation are the spatial domains of a minimum size corresponding to the upper bistability branch, whereas the radiation corresponding to the lower branch forms the background. Dark LSs - vice versa. The mechanisms of the interaction of domain walls, leading to stable LSs, are under investigation, but it is already clear that a nonmonotonic spatial profile of domain boundaries plays a role in LS stabilization [12]. The stronger the spatial oscillations of the „tails" of domain boundaries are, the larger is the stability range of the LSs of the Rosanov type.

In both cases, the spatial modulation of the radiation corresponding to the upper bistability branch stabilizes the bright LS. In the case of Fauve&Thual [6] the parameter range of modulational instability increases with increasing growth exponents for the off-axis modes (A modulational instability is required for LSs of Fauve&Thual type by definition). In the case of Rosanov [7]: the weaker is damping of the spatial oscillations of domain boundaries, the larger is the parameter range of locking of fronts: the more stable are the spatial LSs.

Equivalently the spatial modulation of the radiation corresponding to the lower instability branch supports dark LSs.

Here a general nonlinear optical system with diffusing saturable gain is studied (Section 2), and it is shown, that the diffusion of saturable gain destabilizes predominantly the upper bistability branch. The growth exponents for the off-axis modes are increased due to gain diffusion. Positive growth exponents become even larger, and the parameter range for a modulational instability and for LSs of Fauve&Thual type increases. Negative growth exponents decrease in magnitude, thus spatial oscillations become less damped, and the solitons of Rosanov type become more stable. Finally negative growth exponents may become positive for sufficiently large gain diffusion. Thus one can obtain a transformation of LSs of Rosanov type into LSs of Fauve&Thual type. (This is actually more a transformation of interpretation rather than transformation of the LS itself). In all cases the stability range of the bright LSs is increased with increasing gain diffusion.

These general predictions are illustrated by the example of a laser with saturable absorber in Section III and Section IV.

**II. General case**

In the general case, the mean field equations for an optical system with saturable gain are:

$$\frac{\partial A}{\partial t} = F(A, \nabla^2 A) + DA \tag{1.a}$$

$$\frac{\partial D}{\partial t} = \gamma(D_0 - D - D|A|^2 + d\nabla^2 D) \tag{1.b}$$

here $A(\mathbf{r})$ is the optical field (order parameter), $D(\mathbf{r})$ is the gain field (*e.g.* population inversion). $F(A, \nabla^2 A)$ is a given nonlinear and nonlocal function of order parameter $A(\mathbf{r})$. $d$ is the diffusion coefficient for the saturable gain, $\gamma$ is the relaxation rate of the gain. The complex conjugated equation of (1.a) should be added, if the optical fields are complex (i.e. if diffraction or focusing/defocusing nonlinearities are present in the function $F(A, \nabla^2 A)$).

For simplicity it is assumed below that the gain relaxation is fast ($\gamma = O(1/\varepsilon)$ thus that the gain variable $D$ can be adiabatically eliminated from (1.b). As our numerical calculations show, the main conclusions of the paper are valid even for moderate gain relaxation $\gamma = O(1)$. The adiabatic elimination from (1.b) neglecting gain diffusion ($d = 0$) is straightforward: $D = D_0/(1+|A|^2)$. In the general case ($d \neq 0$) the adiabatic elimination requires inversion of operator $\mathbf{L} = (1+|A|^2 - d\nabla^2)$ since equation (1.b) can be written in the stationary case: $\mathbf{L}D = (1+|A|^2 - d\nabla^2)D = D_0$. The operator inversion can be performed for small diffusion: $d\nabla^2 = O(\varepsilon)$ (all other variables are of $O(1)$) yielding:

$$\mathbf{L}^{-1}D_0 = \frac{D_0}{1+|A|^2} + \frac{D_0}{1+|A|^2} d\nabla^2 \left[\frac{1}{1+|A|^2}\right] + d\nabla\left[\frac{1}{(1+|A|^2)^2}\nabla D_0\right] \tag{2}$$

here the Laplace operator act on the variables to the right. It is easy to verify that $\mathbf{L}^{-1}\mathbf{L}D_0 = D_0(1 + O(\varepsilon^2))$, which confirms the inversion of operator (2) at order of $O(\varepsilon)$.

For spatially homogeneous pump parameter $D_0$ the last r.h.s. term disappears from (2), thus the population inversion becomes:

$$D = \frac{D_0}{1+|A|^2} + \frac{D_0}{1+|A|^2} d\nabla^2\left[\frac{1}{1+|A|^2}\right]. \tag{3}$$

Inserting the (3) into (1.a) one obtains:

$$\frac{\partial A}{\partial t} = F'(A, \nabla^2 A) + \frac{D_0 A}{1+|A|^2} d\nabla^2\left[\frac{1}{1+|A|^2}\right] \tag{4}$$

where $F'(A, \nabla^2 A) = F(A, \nabla^2 A) + D_0 A/(1+|A|^2)$. The last term on the r.h.s. of (4) is due to the diffusion of the saturable gain. On the basis of (4) is investigated below how the gain diffusion affects the stability of both the upper and lower bistability branches.

Linearization of (4) around a homogeneous stationary solution $A_0$: $A = A_0 + a_1 e^{\lambda t}e^{ikr} + a_2^* e^{\lambda t}e^{-ikr}$ leads to:

$$\lambda \mathbf{a} = \mathbf{L}\mathbf{a} + \mathbf{D}\mathbf{a} \tag{5}$$

for a column-vector of perturbation amplitudes: $\mathbf{a} = (a_1, a_2)^T$. $\mathbf{L}$ is the linear evolution matrix generated by the „nondiffusive" part of (4):

$$\mathbf{L} = \begin{bmatrix} \delta F'/\delta a_1 & \delta F'/\delta a_2 \\ \delta F'^*/\delta a_1 & \delta F'^*/\delta a_2 \end{bmatrix} \tag{6}$$

and $\mathbf{D}$ is the perturbation matrix due to gain diffusion:

$$D = \frac{D_0 A_0^2}{(1+A_0^2)^3} dk^2 \begin{bmatrix} 1 & 1 \\ 1 & 1 \end{bmatrix} \qquad (7)$$

here the stationary solution $A_0$ is assumed to be real-valued without loss of generality.

Rewriting (5)-(7) in the new basis: $a_+ = a_1 + a_2$ (the perturbation of the amplitude), and $a_- = a_1 - a_2$ (the perturbation of the phase), one obtains instead of (7):

$$D = \frac{D_0 A_0^2}{(1+A_0^2)^3} dk^2 \begin{bmatrix} 2 & 0 \\ 0 & 0 \end{bmatrix}. \qquad (8)$$

One can draw some conclusions from (8):

1) The sum of the Liapunov exponents is always equal to the trace of the linear evolution matrix. Obviously the perturbation (8) increases the sum of Liapunov exponents by the value: $2dk^2 D_0 A_0^2/(1+A_0^2)^3$. This indicates that overall the gain diffusion works as „anti-diffusion" of the order parameter: the spatial components with high transverse wavenumber (off-axis modes) are amplified due to gain diffusion.

2) If the amplitude and the phase of the order parameter $A$ are decoupled from one another, then the gain diffusion affects only the amplitude perturbations. Therefore the gain diffusion always increases the amplitude modulations, and as a consequence should stabilize spatial solitons.

3) If the amplitude and phase perturbations of the order parameter $A$ are coupled, then the eigenvalues are complex and conjugated: $\lambda_{1,2} = \lambda_{\text{Re}} \pm i\lambda_{\text{Im}}$, thus the sum of eigenvalues is proportional to their real part: $\lambda_1 + \lambda_2 = 2\lambda_{\text{Re}}$. Therefore (8) indicates also the destabilization of coupled amplitude-phase perturbations. The gain diffusion thus increases (or causes) a modulational instability of oscillatory (Hopf) type.

4) The gain diffusion affects predominantly the upper bistability branch: the coefficient of the effective „anti-diffusion" of the order parameter $2dD_0 A_0^2/(1+A_0^2)^3$ is proportional to the field intensity, and is larger for the upper bistability branch than for the lower if the field intensity corresponding to the lower branch is zero or close to zero.

**III. Laser with saturable absorber, analytical results**

We illustrate the above conclusions on the example of a laser with a saturable absorber:

$$F(A, \nabla^2 A) = -A - \frac{\alpha_0 A}{1+|A|^2/I_S} + ia\nabla^2 A + g\nabla^4 A \qquad (9.a)$$

$$F'(A, \nabla^2 A) = \frac{D_0 A}{1+|A|^2} - A - \frac{\alpha_0 A}{1+|A|^2/I_S} + ia\nabla^2 A + g\nabla^4 A \qquad (9.b)$$

here $\alpha_0$ is the coefficient of unsaturated losses, $I_S$ - saturation intensity, $a$ - diffraction coefficient, and $g$ is the field diffusion coefficient (inversely proportional to the width of the gain line). Zero detuning is assumed in (9).

For $g = 0$ one has a purely diffractive case, as studied in [7]. For $a = 0$ one has the purely diffusive case. In optics this purely diffusive case can be realized using self-imaging resonators [13].

First we investigate the purely diffusive case: the amplitude and phase of perturbations are decoupled, thus the equations (4,9) simplify:

$$\partial A/\partial t = \frac{D_0 A}{1+|A|^2} - A - \frac{\alpha_0 A}{1+|A|^2/I_S} - g\nabla^4 A + \frac{D_0 A}{1+|A|^2} d\nabla^2 \left[\frac{1}{1+|A|^2}\right]. \qquad (10)$$

The linear stability analysis of the homogeneous upper branch solution of (10) yields:

$$\lambda = -\frac{2D_0 A_0^2}{\left(1+A_0^2\right)^2} + \frac{2\alpha_0 A_0^2/I_S}{\left(1+A_0^2/I_S\right)^2} - gk^4 + \frac{2D_0 A_0^2}{\left(1+A_0^2\right)^3} dk^2 \qquad (11)$$

for the amplitude perturbations. The phase perturbations are not affected by gain diffusion in this purely diffusive case.

A family of plots of (11) is given in Fig.1.a showing the modulational instability, appearing and growing with an increasing gain diffusion $d$. For sufficiently large gain diffusion the upper branch can be modulationally unstable.

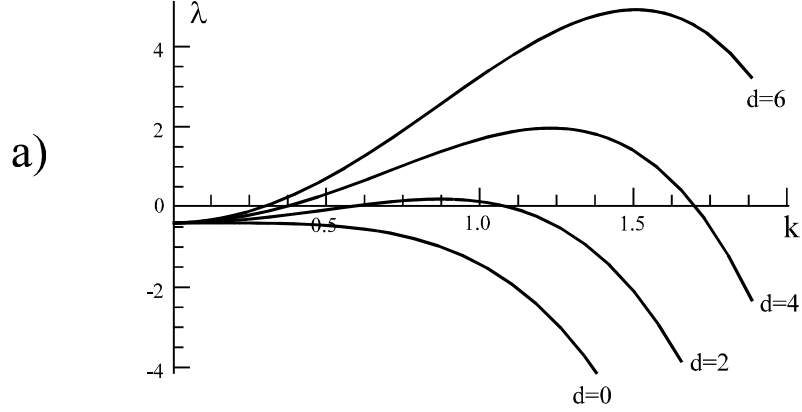

**Fig.1.** Results of linear stability analysis of the upper solution branch: a) for adiabatically eliminated system (10); b) for initial equation system (1,9) b). The Liapunov growth exponents are plotted depending on the transverse wavenumber for different values of gain diffusion $d = 0, 2, 4, 6$. The purely diffusive case: $a = 0, g = 1$. The other parameters: $D_o = 2.8, \gamma = 5, I_s = 0.1, \alpha_o = 5.0$.

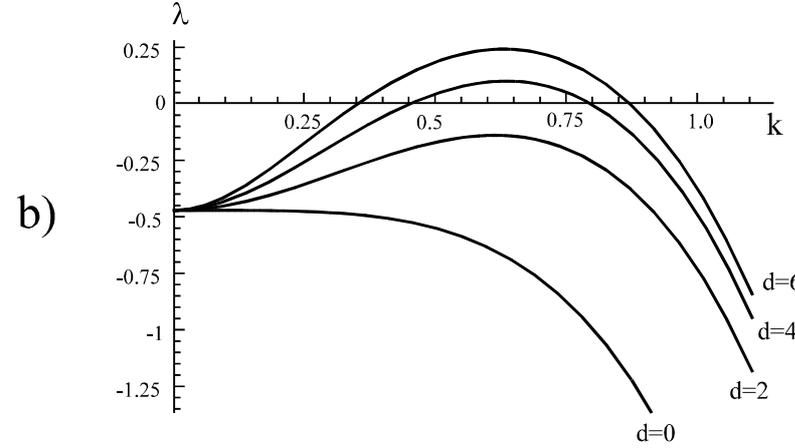

In order to test whether the above procedures of operator inversion and of adiabatic elimination produces correct results, the linear stability analysis of the full problem (equations (1), with (9.a)) was also performed. Figs.1.b shows the results of the stability analysis of the full system. Evidently the instability spectra for small gain diffusion, and small transverse wavenumbers agree well in both cases (the smallness parameter in the adiabatic elimination (2)-(4) is indeed $d\nabla^2 = -dk^2 = O(\varepsilon)$ ). Differences appear for relatively large values of gain diffusion leading to quantitatively different (but not qualitatively different) results.

The stability analysis of the full system shows that in the supercritical case (laser without saturable absorber: $\alpha_0 = 0$) a modulational instability occurs never. Indeed, the expression for upper $\lambda$-exponent branch reads in this case:

$$\lambda = \frac{1}{2}\left[-dk^2 - gk^4 - \gamma D_0 + \sqrt{(dk^2 - gk^4)^2 - 8\gamma(D_0 - 1)}\right]. \qquad (12)$$

The maximal value of the $\lambda$-exponent increases with increasing gain diffusion, and approaches zero from below, for infinitely large gain diffusion. In the case of class-B lasers ($\gamma \ll 1$) the maximal value of the $\lambda$-exponent can approach zero extremely closely, but never becomes positive in this supercritical case.

### IV. Laser with saturable absorber, numerical results

Numerical investigation of the full equation system (equations (1), with (9.a)) were performed. The results are summarized in Fig.2, where the existence ranges of bright and dark LSs are plotted on the plane (gain, diffusion). For comparison, the modulational instability curve (squares), and also the domain equilibrium curve (full circles) is plotted. Typical field profiles corresponding to the different parameter values are plotted in Fig.3.

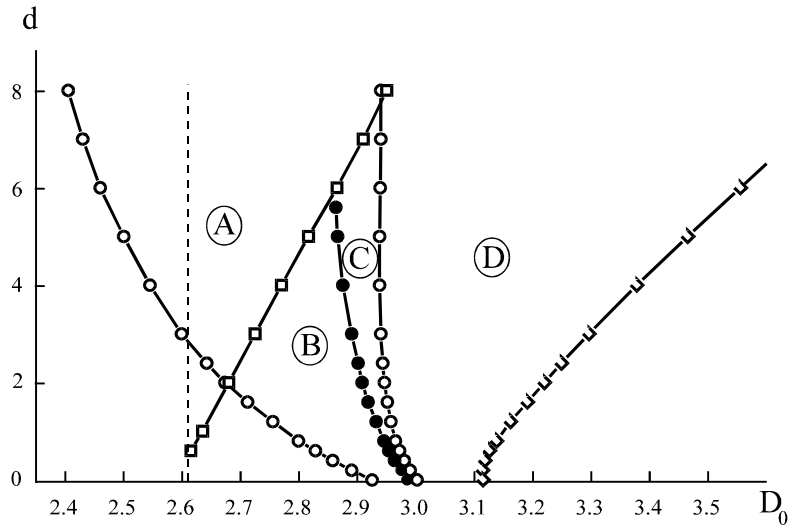

**Fig.2.** Results of numerical analysis of initial equations (1) in case of one spatial dimension: the regions corresponding to the different solutions in the plane of (pump, diffusion). The purely diffusive case. Other parameters are the same as in Fig.1:

Region A: spatial solitons of Fauve&Thual type; Region B: spatial solitons of Rosanov type; Contraction of amplitude domains; Region C: spatial solitons of Rosanov type; Expansion of amplitude domains; Region D+C: dark spatial solitons;

Line with squares (separating regions A and B,C): the modulational instability threshold;
Line with full circles (separating regions B and C): the two phases corresponding to upper and lower solution branch are on equilibrium on this line; The domains neither contract nor expand.
Dashed vertical line: separates monostable and bistable regimes for homogeneous solutions.

In the region A, bright LSs of Fauve&Thual type exist (Fig.3.d). They exist above the modulational instability line in Fig.2 (by definition). In the regions B and C, below the modulational instability line, bright LSs of the Rosanov type are found (Fig.3.c). The region B corresponds to contraction of amplitude domains, and the region C to expansion of the domains, if the domains are sufficiently large. In the case of small domains, the domain boundaries lock and result in stable LSs. As Fig.2 indicates, the locking can occur for contracting domains (region B) and also for expanding domains (region C).

We note here, that the spatial LSs in areas A, B, and C appear visually identical: no abrupt changes of LS parameters are observed when adiabatically crossing the modulational instability threshold.

In general, by increasing the gain diffusion, the existence range of bright LSs increases, and the Rosanov type LSs transform into Fauve&Thual type LSs at the onset of the modulational instability of the upper solution branch.

Fig.3.a and Fig.3.b show equilibrium domains. Between the regions B (domain contraction) and the region C (domain expansion) in Fig.2 the domains are stationary. However, the domain boundaries have no monotonic profile but show spatial oscillations. These oscillations are stronger for larger diffusion of the gain as evident from comparison between Fig.3.a and Fig.3.b. We note that spatial oscillations are much more prominent on the upper bistability branch in correspondence with the predictions above.

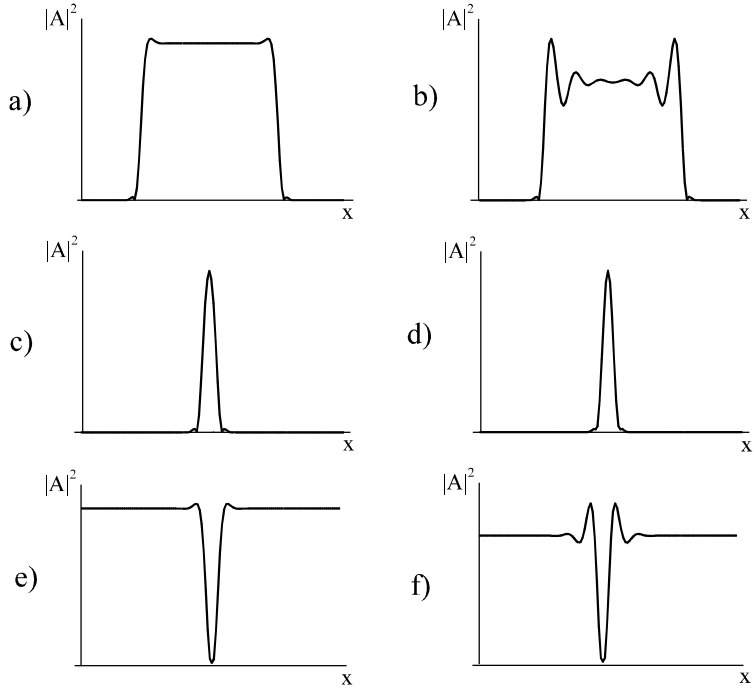

**Fig.3.** Stationary solutions as obtained by numerical integration of initial equations in case of one spatial dimension. The parameters are as in Fig.1,2. a) amplitude domain with weakly nonmonotonic tails: $D_0 = 2.99$, $d = 0$; b) amplitude domain with strongly nonmonotonic tails (close to modulational instability boundary): $D_0 = 2.85$, $d = 5$; c) LS in the region B (of Rosanov type): $D_0 = 2.95$, $d = 0$; d) LS in the region A (of Fouve&Thual type): $D_0 = 2.7$, $d = 5$; e) dark LS in region D for zero gain diffusion (weak spatial modulation): $D_0 = 3.05$, $d = 0$; f) dark LS in region D for strong gain diffusion (strong spatial modulation): $D_0 = 2.15$, $d = 5$.

Dark LSs have also been found numerically. They exist in regions C and D. Interestingly, the enhancement of modulation of the upper branch solution stabilizes dark solitons too. The upper (modulated) solution constitutes the background for dark solitons.

It is usually assumed that the modulation of the solution branch not corresponding to the background stabilizes LSs. For LSs one tends to have a relatively stable background solution branch and a relatively unstable other solution branch. What follows generally from our study of the dark LSs is that the enhancement of the modulation of the *background* also increases the LS stability.

Fig.3.e and Fig.3.f give the numerically calculated field profiles corresponding to dark LSs. Like in the case of large domains, the enhancement of spatial modulation with gain diffusion is also clearly visible in the case of dark LS.

In the not purely diffusive case the mathematical expressions following from the linear stability analysis are not so transparent. The corresponding plots are given in Fig.4. In general, a relatively small field diffraction does not bring about qualitative changes: an enhancement of the modulational instability (Fig.4.a) is observed due to gain diffusion, like in the purely diffusive case studied above. The additional feature compared with the purely diffusive case is the locking between amplitude and phase instabilities in a certain band of transverse wavenumbers (where the amplitude and phase $\lambda$ - branches are sufficiently close). As our numerical calculations show, the enhancement of the LS stability range due to gain diffusion is similar to the purely diffusive case.

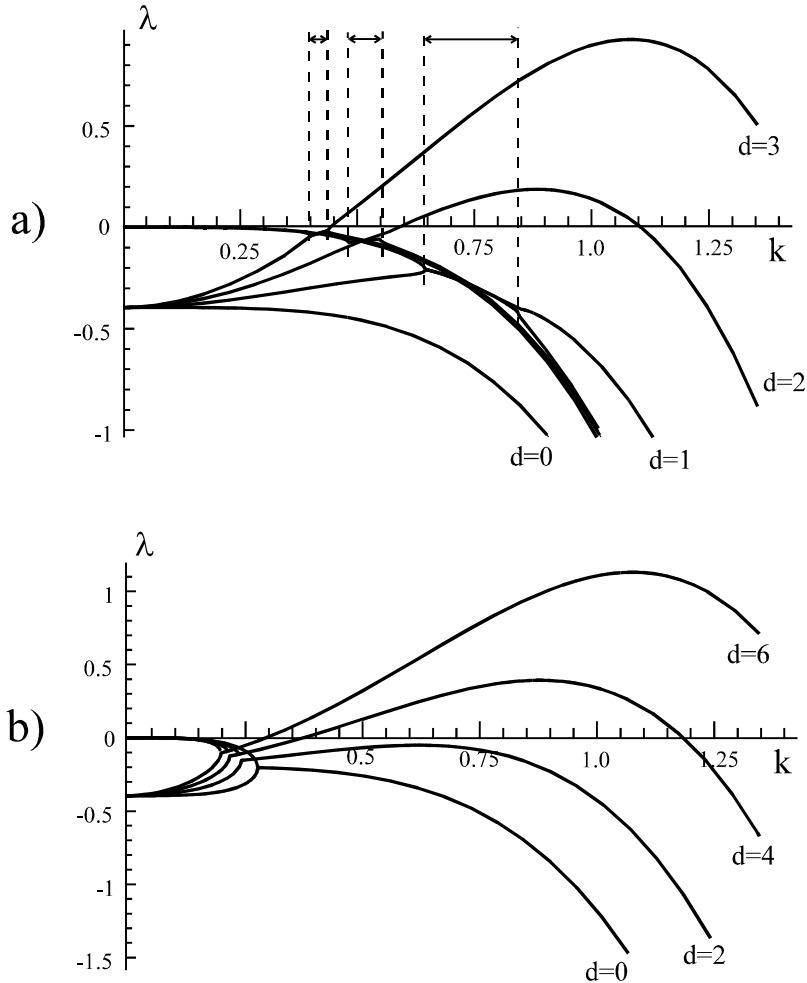

**Fig.4.** Results of linear stability analysis of upper solution branch for adiabatically eliminated system for not purely diffusive case. Parameters as in Figs.1-3, except for diffraction coefficient.

a) weak diffraction: $a = 0.25$; for nonzero gain diffusion small region of locking between amplitude and phase perturbations appear (indicated by vertical dashed lines).

b) strong diffraction $a = 2.5$; the amplitude and phase perturbations are locked everywhere except for relatively small transverse wavenumbers $k$.

For strong diffraction (Fig.4.b), the locking between amplitude and phase perturbations is stronger. As a consequence a nonstationary modulational instability is predicted (instability of Hopf type). One, therefore, may expect oscillatory LSs in the case of strong diffraction. The oscillatory LS were indeed found in a variety of nonlinear optical systems [14]. Detailed study of the patterns and of spatial LSs in this limit of strong diffraction is beyond the present paper.

### V. Conclusions

1. The diffusion of saturable gain enhances the growth of the off-axis field components. Overall the gain diffusion results in „anti-diffusion" of the order parameter. As a result the gain diffusion can enhance and/or cause a modulational instability.

The physical mechanism of modulational instability due to pump diffraction is similar to that of „local activation and lateral inhibition" studied by Turing [15]. In our case the optical field can be identified with the activator, and the gain with the inhibitor. Then the modulational instability (and pattern formation in general) appears due to an interplay between diffusions (or diffractions) of optical field (activator) and gain media (inhibitor). This interpretation has been investigated in detail in [16].

2. The solution corresponding to the upper bistability branch is predominantly affected by gain diffusion. The solution corresponding to the lower branch is almost unaffected when its amplitude is sufficiently smaller than that of the upper branch.

3. As the result, the gain diffusion increases the stability range of the LSs of both types: In the case of Fauve&Thual type LSs, where the LSs apriori require modulationally unstable solutions, the enhancement of the modulation instability obviously increases the existence range of LSs. In the case of LSs of Rosanov type, where the stabilization is due to nonmonotonically decaying fronts of domains, the increase of the gain diffusion results in the increase of spatial oscillations, thus the existence range of LS increases correspondingly.

4. The transition between the LSs of the above types is smooth (no singular behaviour appears at the boundary between the two types of LSs of Rosanov and Fauve&Thual. This suggests, that distinguishing the two types of LSs is only a matter of interpretation. In essence the LSs of both types are similar, as they convert one into another smoothly.

5. As the analysis of bright LSs shows, the enhancement of modulation of the upper state (different from that corresponding to the background solution) increases the stability of a LS. However, the analysis of the dark LSs shows, that also the enhancement of modulation of the background state stabilizes the LSs.

6. In the diffractive case the pump diffusion can enhance or cause not only the stationary modulational instabilities, but also the nonstationary ones (instabilities of Hopf type). The behaviour of LSs in the strongly diffractive regime, characterized by a nonstationary (Hopf) modulational instability was not studied in detail in the present paper.

7. The above conclusions concern diffusion of *gain*. If instead of the diffusing gain in (1.b) one has diffusing losses (*e.g.* if the amplifying media in the resonator is not pumped: $D_0 < 0$) then the conclusions are opposite. The diffusion of the losses then stabilize the upper bistability branch and reduce the stability range of the spatial LSs, as follows from (10) and (11) with $D_0 < 0$.

This case seemingly corresponds to the investigations in [10,11], where semiconductor lasers were investigated in different regimes: i) in the amplifying regime both above and below the generation threshold, ii) as a driven nonlinear resonator. The free carriers in the case ii) cause saturating losses, rather than saturating gain. Therefore it is plausible that the diffusion of free carriers reduces the modulational instability and plays a destructive role for the spatial LS. Even in the amplifying regime the gain in semiconductor lasers is complex-valued (the parameter $D_0$ in (1.b)). Therefore, as follows from (10) the diffusion of free carriers results in Laplace operators with complex coefficients, and the further analysis on the basis of (10) is complex.

**Acknowledgment**

Discussions with C.O.Weiss, V.Taranenko, V.J.Sanchez-Morcillo, D.Michaelis and M.Brambilla are acknowledged.